\documentclass[longauth]{aa}
\usepackage{graphicx}
\usepackage{siunitx}
\usepackage{comment}
\usepackage{caption}  
\usepackage{txfonts}
\usepackage{hyperref}
\newcommand{  \hi       }{\ifmmode {\rm H}\,\textsc{i}\ \else H\,\textsc{i}\,\fi}

\newcommand \BB {$^{\text{3D}}$Barolo}
\newcommand{\sample}{NGC~1512, NGC~4535, and NGC~7496}

\newcommand{\orcid}[1]{\href{https://orcid.org/#1}{\includegraphics[width=10pt]{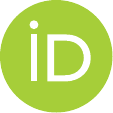}}}

\DeclareSIUnit\parsec{pc} 
\DeclareSIUnit\magnitudes{mag} 
\DeclareSIUnit\arcsec{arcsec}

\newcommand{\Bonn}{Argelander-Institut f\"ur Astronomie, Universit\"at Bonn, Auf dem H\"ugel 71, 53121 Bonn, Germany}

\newcommand{\NRAO}{National Radio Astronomy Observatory, 520 Edgemont Road, Charlottesville, VA 22903, USA}

\newcommand{\OSU}{Department of Astronomy, The Ohio State University, 140 West 18th Ave, Columbus, OH 43210, USA}

\newcommand{\CCAP}{Center for Cosmology and Astroparticle Physics, 191 West Woodruff Avenue, Columbus, OH 43210, USA}

\newcommand{\astron}{Netherlands Institute for Radio Astronomy (ASTRON),  Oude Hoogeveensedijk 4, 7991 PD Dwingeloo, Netherlands}

\newcommand{\kapeyn}{Kapteyn Astronomical Institute, University of Groningen, PO Box 800, 9700 AV Groningen, The Netherlands}

\newcommand{\UCT}{Department of Astronomy, University of Cape Town, Private Bag X3, 7701 Rondebosch, South Africa}

\begin{document}

   \title{Neutral atomic and molecular gas dynamics in the nearby spiral galaxies NGC 1512, NGC 4535, and NGC 7496}


   \author{Sebastian Laudage
          \inst{1,}\thanks{\email{laudage@uni-bonn.de}}
          \orcid{0009-0002-4351-7301}
          \and
          Cosima Eibensteiner\inst{1, 2, }\thanks{Jansky Fellow of the National Radio Astronomy Observatory, \email{ceibenst@nrao.edu}} \orcid{0000-0002-1185-2810}
          \and
          Frank Bigiel\inst{1}\orcid{0000-0003-0166-9745}
          \and
          Adam K. Leroy\inst{3, 4}\orcid{0000-0002-2545-1700}
          \and
          Sharon Meidt\inst{5}\orcid{0000-0002-6118-4048}
          \and\\
          Eva Schinnerer\inst{6}\orcid{0000-0002-3933-7677}
          \and
          W.J.G. de Blok\inst{7,8,9}\orcid{0000-0001-8957-4518}
          \and
          Miguele Querejeta\inst{10}\orcid{0000-0002-0472-1011}
          \and
          Sophia Stuber\inst{6}\orcid{0000-0002-9333-387X}
          \and
          Dario Colombo \inst{1}\orcid{0000-0001-6498-2945}
          \and\\
          Erik Rosolowsky\inst{11}\orcid{0000-0002-5204-2259}
          \and
          D.J. Pisano\inst{9}\orcid{0000-0001-7996-7860}
          \and
          Dyas Utomo\inst{3}\orcid{0000-0003-4161-2639}
          \and
          Rebecca C. Levy\inst{12}\orcid{0000-0003-2508-2586}
          \and
          Ralf Klessen\inst{13}\orcid{0000-0002-0560-3172}
          \and
          Yixian Cao\inst{14}\orcid{0000-0001-5301-1326}
          \and\\
          Eric W. Koch\inst{15}\orcid{0000-0001-9605-780X}
          \and
          Sushma Kurapati\inst{9}\orcid{0000-0001-6615-5492}
          \and
          Patricia Sanchez-Blazquez\inst{16}\orcid{0000-0003-0651-0098}
          \and
          Justus Neumann\inst{6}\orcid{0000-0002-3289-8914}
          \and
          Lukas Neumann\inst{1}\orcid{0000-0001-9793-6400}
          \and
          Hsi-An Pan\inst{15}\orcid{0000-0002-1370-6964} 
          \and
          Thomas G. Williams\inst{18}\orcid{0000-0002-0012-2142}
          }

   \institute{\Bonn\
              \and 
              \NRAO\
              \and
              \OSU\
              \and
              \CCAP\
              \and
              Sterrenkundig Observatorium, Universiteit Gent, Krijgslaan 281 S9, B-9000 Gent, Belgium\
              \and
              Max Planck Institute for Astronomy, Königstuhl 17, 69117 Heidelberg, Germany\
              \and 
              \astron\
              \and
              \kapeyn\
              \and
              \UCT\
              \and 
   Observatorio Astron{\'o}mico Nacional (IGN), C/Alfonso XII 3, Madrid E-28014, Spain
   \and
   University of Alberta, Edmonton, Alberta, T6G 2E1, Canada
   \and
   Steward Observatory, University of Arizona, Tucson, AZ 85721, USA
   \and
   Universit\"{a}t Heidelberg, Zentrum f\"{u}r Astronomie, Albert-Ueberle-Str. 2, 69120 Heidelberg, Germany
   \and
   Max-Planck-Institut f\"{u}r extraterrestrische Physik, Giessenbachstra{\ss}e 1, D-85748 Garching, Germany
   \and
   Center for Astrophysics $\mid$ Harvard \& Smithsonian, 60 Garden St., 02138 Cambridge, MA, USA
   \and
   Departamento de Fisica de la Tierra y Astrofisica, Facultad de CC Fisicas, Universidad Complutense de Madrid, 28040, Madrid, Spain
   \and
   Department of Physics, Tamkang University, No.151, Yingzhuan Road, Tamsui District, New Taipei City 251301, Taiwan
   \and
   Sub-department of Astrophysics, Department of Physics, University of Oxford, Keble Road, Oxford OX1 3RH, UK} 
   
   \date{Received ; accepted}

   \abstract{Neutral atomic gas ($\rm H\,\textsc{i}$) effectively traces galactic dynamics across mid to large galactocentric radii. However, its limitations in observing small-scale changes within the central few kiloparsecs, coupled with the often observed \hi deficit in galactic centers, necessitates the use of molecular gas emission as a preferred tracer in these regions. Understanding the dynamics of both neutral atomic and molecular gas is crucial for a more complete understanding of how galaxies evolve, funnel gas from the outer disk into their central parts, and eventually form stars. In this work we aim to quantify the dynamics of both, the neutral atomic and molecular gas, in the nearby spiral galaxies NGC~1512, NGC~4535, and NGC~7496 using new MeerKAT \hi\ observations together with ALMA CO (2-1) observations from the PHANGS collaboration. We use the analysis tool \BB\ to fit tilted ring models to the \hi\ and CO observations. A combined approach of using the \hi\ to constrain the true disk orientation parameters before applying these to the CO datasets is tested. This paper sets expectations for the results of the upcoming high-resolution \hi coverage of many galaxies in the PHANGS-ALMA sample using MeerKAT or VLA, to establish a robust methodology for characterizing galaxy orientations and deriving dynamics from combing new \hi\ with existing CO data.}


   \titlerunning{Kinematic analysis of nearby spiral galaxies NGC 1512, NGC 4535, and NGC 7496}
   \authorrunning{Laudage et al.}
   \maketitle
   
\section{Introduction}
Different emitting constituents (neutral gas, ionized gas, molecular gas, or stars) of star-forming disk galaxies and their individual rotation curves (RCs) can be used to trace galactic dynamics. Each of these galactic components offers a unique perspective on the dynamics caused by their distinct and characteristic radial and vertical distribution \citep{2014ApJ...784....4C,2020MNRAS.496.1845M}. The ubiquitous neutral atomic gas ($\rm H\,\textsc{i}$) is typically the most extended component in galaxies reaching out to $2-4\times \text{r}_{25}$ (\citealt{2016MNRAS.460.2143W}; where r$_{25}$ is the radial position of the isophote at 25~mag~arcsec$^{-2}$).

\hi\ kinematics have long been used to trace the galactic potential and its division into baryonic and dark matter mass components \citep[e.g.][]{2008AJ....136.2648D,2002A&A...385..816D,2007MNRAS.375..199G,2016AJ....152..157L}. The extended atomic gas disks in particular, provide a key view on the dark matter halos of galaxies \citep[e.g.][]{1981AJ.....86.1825B, 2001ARA&A..39..137S} and the accretion of new material that supplies the fuel for star formation \citep[see e.g.][]{2016MNRAS.457.2642S, 2021ApJ...923..220D, Cosi}.

To trace small-scale kinematics in the centers of galaxies that are rich in dynamical structure (e.g. having bars or rings), CO becomes preferred over $\rm H\,\textsc{i}$. This is because the ISM tends to be molecular-dominated towards the center and CO observations can achieve better resolutions  \citep{2012ApJ...756..183B,2018ApJ...860...92L,2022ApJ...934..173S, 2008AJ....136.2563W}. One of the issues often faced when modeling the kinematics in the centers is that the gravitational potential and gas kinematics show signs of non-axisymmetry (i.e due to stellar bars and spiral arms), which can make it difficult to accurately determine the disk orientation and RC \citep{2020ApJ...897..122L, 2006ApJ...646..213K} at these radii. Additionally, the molecular gas traced by CO tends to be more clumpy in comparison to \hi, making CO velocity fields sometimes sparsely sampled, presenting a further challenge for kinematic modeling \citep{2020ApJ...897..122L}.

Given these difficulties, a solution would be to use the \hi velocity field sampling further out in the disk to constrain the true disk orientation, thereby improving the accuracy in RC determination from either the CO or \hi information. This paper examines the validity and usefulness of this approach, studying an initial small sample of galaxies with both existing CO and new \hi data.  The hypothesis is that \hi can provide a good probe of the gas disk's intrinsic orientation, as long as additional factors that affect \hi and CO are taken into consideration on a galaxy-by-galaxy basis.  These factors include the fact that outer \hi disks of galaxies often exhibit genuine warps and twists as well as spiral arms \citep[e.g.][]{2004ApJ...605..183W, 2008AJ....136.2648D,2016MNRAS.457.2642S,2008AJ....136.2720T}. \hi kinematics may also trace radial inflows, as newly accreted material in the outskirts of the disk makes its way inward \citep[e.g.][]{2004ApJ...605..183W,2021ApJ...923..220D, Cosi}. All of these factors can impact the derivation of the disk parameters. Consequently, both the orientation must be allowed to vary when modeling \hi velocity fields, and the validity of that orientation in the inner CO disk is not guaranteed.

In this paper, we perform a detailed modeling of the CO and \hi observations to determine where their RCs agree and under which circumstances they differ. Our initial target consists of three galaxies, but it should be sufficient for reaching our goal of determining a best-practice recommendation for fitting the disk orientation and RC when both CO and \hi observations are available in future works. 

For our dynamic analysis and comparison of atomic and molecular rotation curves, we use observations from the PHANGS collaboration (Physics at High Angular resolution in Nearby GalaxieS\footnote{\url{www.phangs.org}}) \citep{2019Msngr.177...36S}. The PHANGS-ALMA survey presents 90 nearby ($d\leq\SI{20}{\mega\parsec}$ ) galaxies targeting CO J=2$\rightarrow$1 emission \citep{2021ApJS..257...43L}. 
By using the excellent capabilities of the MeerKAT telescope it is possible to complement these data with a unique view in \hi. \citet{Eibensteiner2024} present a refined view of the \hi in eight nearby star-forming galaxies observed with the MeerKAT telescope. Three of these galaxies are part of the PHANGS-MeerKAT sample (cycle0; PI: D. Utomo) which is growing in the near future (cycle1; PI: D.J. Pisano). The focus of this paper is on the first three galaxies in the PHANGS-MeerKAT sample: NGC~1512, NGC~4535, and NGC~7496.

In addition to pure circular velocities, we also analyze radial velocities in \hi of the three galaxies. Studying radial motions of matter in galaxies contributes to the explanation of their current star formation rates and evolution \citep{2004ApJ...605..183W, 2007ApJ...664..204S}. Many galaxies, including the Milky Way, have been forming stars at a nearly constant or slightly decreasing rate throughout cosmic time \citep{ 2011ApJ...730L..13B, 2007MNRAS.378.1550P}. This raises the question of why the star-forming gas is not yet depleted. A prominent explanation for the current star formation rates (SFRs) is radially transported neutral atomic gas \citep[see e.g.][]{2016MNRAS.457.2642S}. We analyze radial mass flow rates even though our sample size is currently not large enough to formulate conclusions on this topic. However, with upcoming \hi observations the presented methodology in this paper can be applied to a much larger sample. 

   \begin{figure*}[h!]
   \centering
   \includegraphics[width=0.85\textwidth]{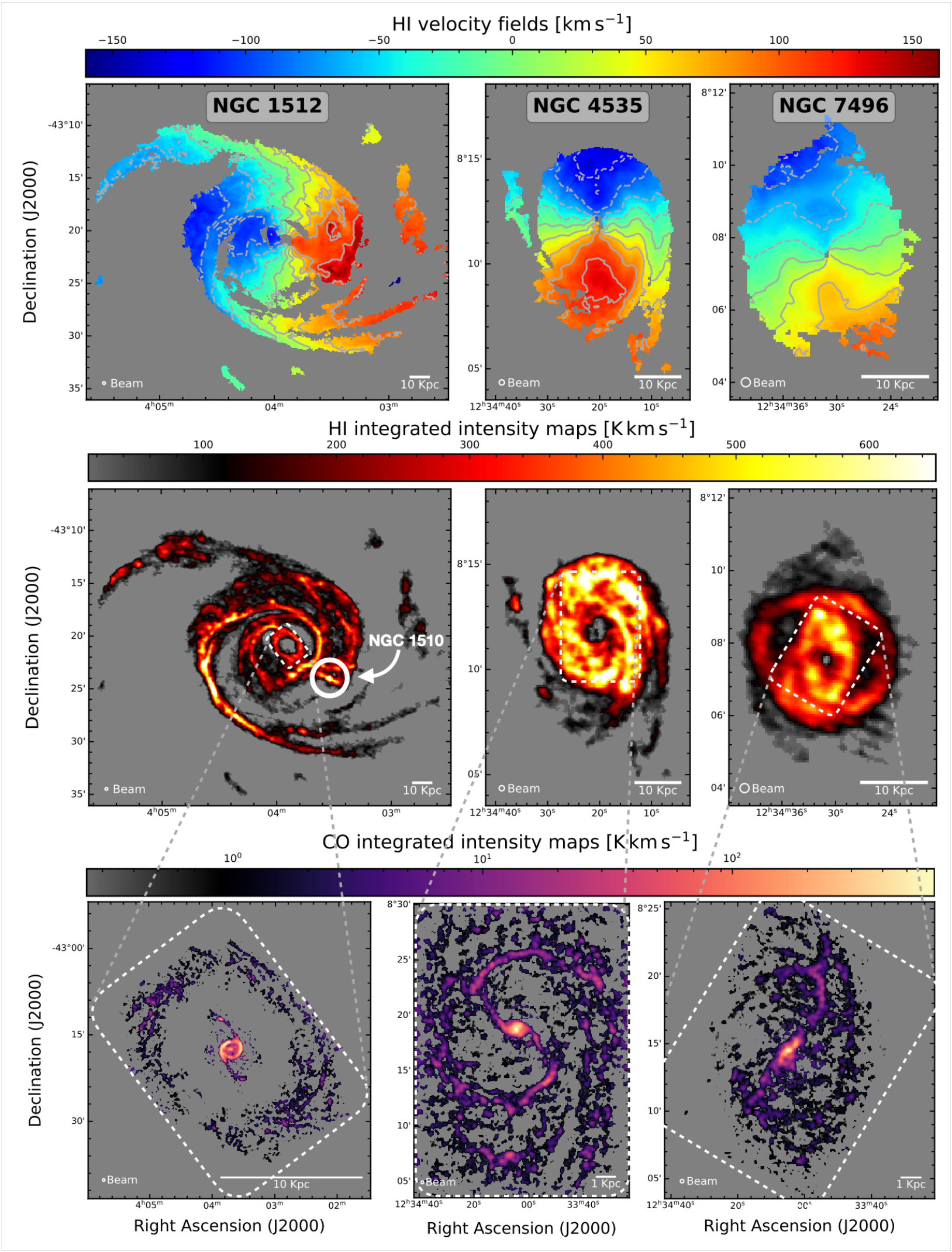}
   \caption{\textit{Top row:} Moment 1 maps from PHANGS MeerKAT \hi\ observations of NGC 1512, NGC 4535, and NGC 7496 built with the strict masking procedure described in Section \ref{sec:Methods}.  The maps are overplotted with isovelocity-lines (grey) in intervals of \SI{30}{\kilo\meter\per\second}. \textit{Middle row:} Integrated intensity or Moment 0 maps from the MeerKAT \hi\ observations built with the same masking procedure. The location of the companion galaxy NGC 1510 is indicated by the white circle. \textit{Bottom row:} Integrated intensity or Moment 0 maps of the three galaxies built from CO observations taken from PHANGS-ALMA. The extent of the CO maps is indicated in comparison to the \hi\ maps.}
              \label{fig:moment1_HI}%
   \end{figure*}

We describe our observations in Section \ref{sec:observations}. Section \ref{sec:Methods} outlines our methodology for implementing the tilted ring model. We present our results that encompass the best-fit parameters and rotation curves, in Section \ref{sec:results}. Section \ref{discussion}, is dedicated to the implications and trustworthiness of our models and we further quantify radial motions in our models.

\begin{table*}[ht]
\caption{Physical Properties of the MeerKAT galaxies NGC 1512, NGC 4535 and NGC 7496. The columns from left to right contain the morphological type, central right ascension (RA) and declination (Dec.), the distance, the systemic velocity, the stellar effective radius, and the bar radius. The values are taken from: (1) \citep{Makarov2014}, (2) \citep{2015ApJS..219....4S}, (3) distanced are curated by \citep{Anand2021}(with original references for NGC 1512: \citet{Scheuermann2022}; for NGC 4535: \citet{Freedman2001}, and for NGC 7496: \citet{Shaya2017} and \citet{Kourkchi2020})}, (4) \citep{2020ApJ...897..122L},(5) \citep{2021ApJS..257...43L}, (6) \citep{2021A&A...656A.133Q}.  
\label{table:Properties}      
\centering     
\begin{tabular}{l c c c c c c c}    
\hline\hline
Galaxy ID & Morphology & $\alpha_{\text{J2000}}$  & $\delta_{\text{J2000}}$ & Distance (\SI{}{\mega\parsec}) & V$_{\text{sys}}$ (\SI{}{\kilo\meter\per\second})& R$_{\text{eff}}$ (\SI{}{\kilo\parsec}) & R$_{\text{bar}}$ (\SI{}{\kilo\parsec})\\ 
(Reference) & (1) & (2) & (2) & (3) & (4) & (5) & (6) \\
\hline
NGC 1512 & SB(r)ab & $4^{\text{h}}03^{\text{m}}54.1^{\text{s}}$ & $-\SI{43}{\degree}\SI{20}{\arcminute}\SI{55}{\arcsecond}$  & $17.92\pm0.88$ & $871.4\pm5$ & 4.8 & 6.5\\ 
NGC 4535 & SAB(s)c & $12^{\text{h}}34^{\text{m}}20.31^{\text{s}}$ & $+\SI{8}{\degree}\SI{11}{\arcminute}\SI{56}{\arcsecond}$  & $15.77\pm0.37$ & $1953.6\pm5$ & 6.3 & 2.9\\
NGC 7496 & SBb & $23^{\text{h}}09^{\text{m}}47.3^{\text{s}}$ & $-\SI{43}{\degree}\SI{25}{\arcminute}\SI{40}{\arcsecond}$ & $18.72\pm2.82$ & $1639.2\pm5$ & 3.8 & 3.4\\
\hline  
\end{tabular}
\end{table*}

\section{Sample and Observations}
\label{sec:observations}

\subsection{Sample}
The galaxies \sample\ are all nearby (d$\lesssim$\,20 Mpc), star-forming, barred spiral galaxies (\autoref{table:Properties}). Targets were selected from the parent sample of galaxies that have PHANGS-ALMA CO observations and where MeerKAT has an opportunity to substantially improve on the quality of data available in the literature because of its excellent $uv$ coverage, sensitivity and resolution capabilities. Previous \hi studies of nearby galaxies that are part of the PHANGS-ALMA sample \citep[e.g][]{2018MNRAS.478.1611K, 2008AJ....136.2648D, 2010MNRAS.402.2403M, 2024arXiv240401774D,2009AJ....138.1741C} have led to a detailed view of the atomic gas distribution already. 
NGC 1512 for example was covered with the Australia Telescope Compact Array (ATCA) \citep{2009MNRAS.400.1749K} and additionally observed at a coarser resolution with MeerKAT \citep{2023arXiv230902076E}. NGC 4535 has been mapped as part of the  Imaging survey of Virgo
galaxies in Atomic gas (VIVA) \citep{2009AJ....138.1741C} before.

Moment maps in \hi\ and CO and the moment 1 map in \hi\ are presented in \autoref{fig:moment1_HI}. The most unique appearance is seen in NGC 1512. It has a double-ring structure, a nuclear ring around the galactic center, and an additional ring further out in the main disk. The galaxy presents a very extended disk in \hi\ (going out to $\sim\SI{120}{\arcsecond}$) with the dwarf galaxy NGC 1510 located about \SI{26.5}{\kilo\parsec} away (assuming the distance presented in table \autoref{table:Properties}). A clear peak in \hi emissions is seen at the position of NGC 1510 from our integrated intensity map (compare \autoref{fig:moment1_HI}. The galaxy pair is believed to be strongly interacting because of the proximity, creating an asymmetric distribution of \hi across the disk of NGC 1512 \citep{2009MNRAS.400.1749K}. A similar but, not as strong unwinding of spiral arms can be seen in NGC 4535. Such unwinding of spiral arms is typical for galaxies influenced by ram pressure inside galaxy clusters \citep{2021MNRAS.500.1285B}. NGC 4535 is part of the Virgo Cluster and the peculiar structure at large radii might be caused by this effect.

\subsection{MeerKAT 21cm observations}

The \hi observations used in this work were taken during the first observing cycle (Cycle 0) of the new MeerKAT (Meer Karoo Array Telescope) radio interferometer (PI: D. Utomo). It is the most sensitive centimeter-wavelength interferometer in the southern hemisphere. The special design of the MeerKAT antennas, results in low system temperatures and an antenna gain of $\sim$2.8 K/Jy \citep{2016mks..confE...1J}. These properties make it ideally suited to detect low column density \hi emission \citep[see e.g.][]{2020Ap&SS.365..118K,2016mks..confE...7D}.

The galaxies (see \autoref{table:Properties}) were each observed for 6 hours in a frequency range from \SI{1.411}{\giga\hertz} to \SI{1.414}{\giga\hertz}. The observations were imaged to a resulting angular resolution of \SI{15}{\arcsecond} and a spectral resolution of \SI{5.5}{\kilo\meter\per\second} per channel \citep[see][for more details]{Eibensteiner2024}.
We used the strict masking procedure described in \citet{2021ApJS..257...43L} that was initially used for the PHANGS-ALMA CO(2-1) observations. This high-confidence masking includes only voxels with a high signal-to-noise ratio and few or no noise-dominated sight lines. This is achieved by creating a core mask that includes all voxels with a signal-to-noise (S/N) larger than 4 in two consecutive velocity channels and a lower S/N outer mask that includes voxels with a S/N larger than 2 in two consecutive velocity channels. The strict mask then consists of all contiguous regions in the outer mask that contain pixels from the high-significance inner mask. 

\subsection{PHANGS-ALMA CO observations}
To trace the molecular component we used observations from the PHANGS-ALMA survey \citep{2021ApJS..257...43L}. The survey maps the CO $\text{J}=2\rightarrow1$ line emission in these galaxies at an angular resolution of ${\sim}1''$ which corresponds to linear scales of ${\sim}\SI{100}{\parsec}$. The spectral resolution is \SI{2.5}{\kilo\meter\per\second}. For CO we used a different masking procedure than for the \hi described above. As CO emission is generally less smoothly distributed, and good spatial coverage is needed for reliable results, we used another product offered by the PHANGS-ALMA pipeline, the "moment1 wprior" map. These maps include all moment 1 values calculated from the strict mask (procedure described above), but add line-of-sights that are in agreement with a prior guess (the prior is created from a strictly masked \SI{15}{\arcsecond} resolution cube). This process can result in increased spatial coverage of the moment-1 map at high resolution compared to the strictly masked moment-1 maps, while still being reliable at rejecting severe outliers in the velocity space \citep[see][for more details]{2021ApJS..257...43L}. 

\section{Methods}
\label{sec:Methods}
RCs are commonly derived by fitting tilted ring models to the velocity field of the line emission and extracting the rotational velocity parameter from the best-fit model. Several available software packages exist that support this approach such as ROTCUR \citep{1987hrcs.book.....B} or DISKFIT \citep{2007ApJ...664..204S}. These two-dimensional approaches provide reliable RCs for highly resolved data but have the disadvantage of being dependent on assumptions made during the extraction of the velocity fields. 

The beam-smearing effect is another severe problem in deriving kinematics from the velocity field \citep{1981AJ.....86.1825B}. The effect is caused by the finite size of a telescope's beam which causes es the line emission to be smeared in the adjacent regions. As a typical result, the rising part of the extracted rotation curve can be underestimated. 

A solution to take both mentioned effects into account is to use a 3-dimensional tilted ring modeling algorithm such as TiRiFiC \citep{2007A&A...468..731J} or \BB~ \citep{2015MNRAS.451.3021D}. In this paper, we compute rotation curves using \BB\ which fits tilted ring models to emission line data cubes in three dimensions. It does not depend on the methodology of the velocity field derivation and takes beam smearing into account by introducing the instrumental effect to the model during the convolution. 
\subsection{Tilted Ring Model}
   \begin{figure*}[ht]
   \centering
   \includegraphics[width=1\textwidth]{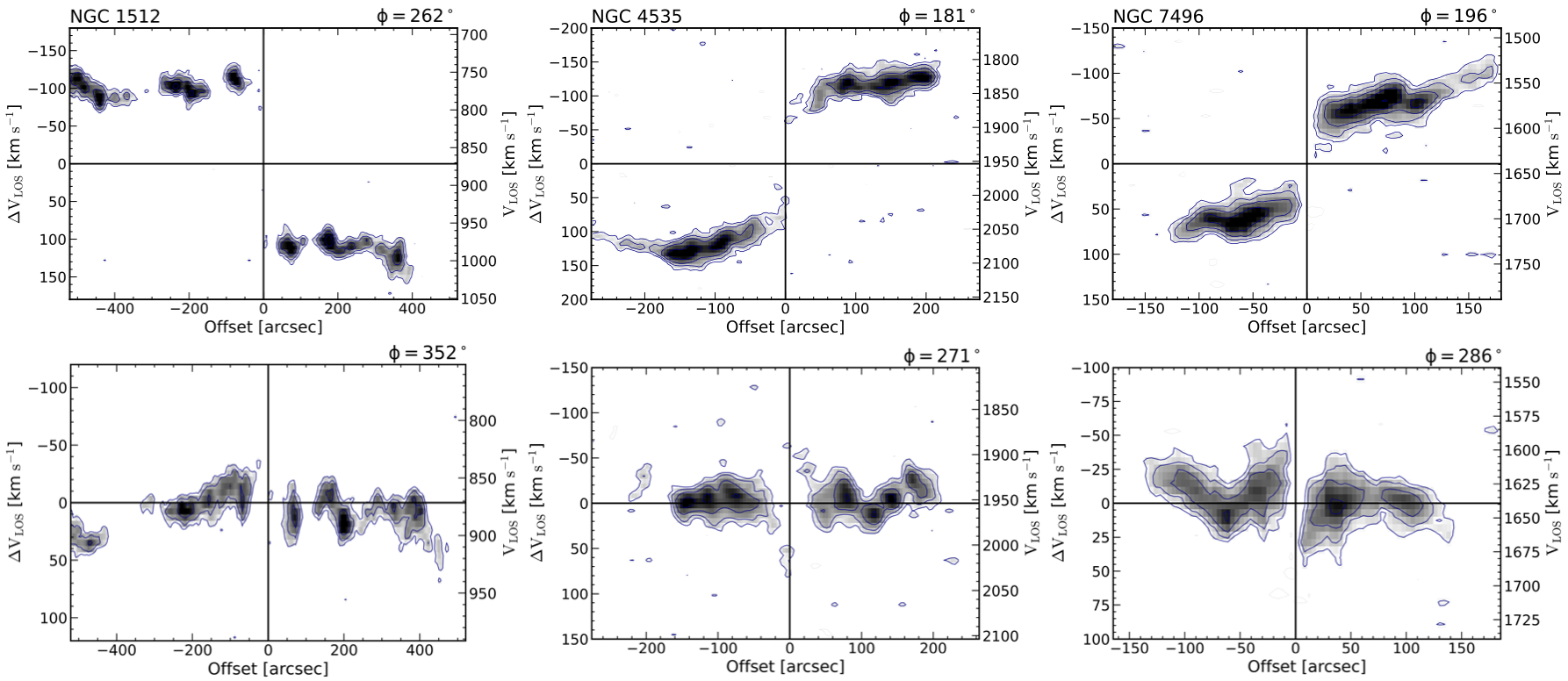}
   \caption{\centering \textit{Top row:} Line-of-sight velocities along the major axis position from the \hi observations of NGC 1512 (left), NGC 4535 (middle), and NGC 7496 (right). \textit{Bottom row:} Line-of-sight velocities along the minor axis. The x-axis represents the offset of the galactic center and the y-axis represents the observed velocities with (left) and without (right) subtracting the systemic velocity of the galaxy.}
              \label{fig:PVs_MAJOR}%
   \end{figure*}
\BB\ shares the basic scheme adopted by any tilted-ring model, modeling the geometry and dynamics of a galaxy by subdividing the galactic disk radially into several circular rings. Each ring is described by a set of parameters that (at first) do not depend on parameters from other rings. The ring parameters are connected to line-of-sight velocities via a harmonic expansion:
\begin{align}
\label{vloss}
V_{\text{los}}(R, \theta)=V_{\text{sys}}+V_{\text{rot}}(R)\,\text{cos}\,\theta\, \text{sin}\,i+V_{\text{rad}}(R)\,\text{sin}\,\theta\, \text{sin}\,i
\end{align}
with $i$ being the inclination angle and $\theta$ the azimuthal angle measured inside the galactic disk \citep{2017MNRAS.466.4159I}. The ring radius $R$ is defined as the median distance of each ring to the center. 
The orientation of the projection of the inclined galactic disk is described by the position angle $\phi$. It is an azimuthal angle between the north direction and the receding major axis of the projection on the celestial sphere measured eastwards. The connection between $\theta$ and $\phi$ is given by the following relations:
\begin{align}
    \label{cos}
    \cos{\theta}=\frac{-(x-x_0)\,\sin{\phi}+(y-y_0)\,\cos{\phi}}{R}\\
    \sin{\theta}=\frac{-(x-x_0)\,\cos{\phi}+(y-y_0)\,\sin{\phi}}{R\,\cos{i}}
\end{align}
with $x_0$ and $y_0$ describing the coordinates of the center of the rings. This description yields line-of-sight velocities based on a set of parameters for an arbitrarily chosen coordinate in the disk. With the equation, it is possible to fit these parameters to an observed velocity field.

\subsection{Initial Parameters and Assumptions with \BB}

\label{sec:initial parameters}
Finding the best initial parameters for the fitting procedure is one of the most important steps when working with algorithms like \BB. The two-dimensional tilted ring model already involves seven fitting parameters (see Equation \ref{vloss} and Equation \ref{cos}), namely $V_{\text{sys}}, x_0, y_0, i, \phi, V_{\text{rad}}$, and $V_{\text{rot}}$. In the 3D version, three additional parameters are needed to apply a model directly to the emission line data cube, which are $z_0$ (the thickness of the disk), $\Sigma_{\text{gas}}$ (gas surface density), and $\sigma_{\text{gas}}$ (the velocity dispersion in the gas disk). The used initial parameters were taken from previous works on these galaxies and are listed with sources in \autoref{tab:Initial Parameters}.

\begin{table}[ht]
\caption{A total of 9 parameters are used during the fitting: The systemic velocity $V_{\text{sys}}$, the coordinates of the galactic center given in right ascension and declination, the rotational velocity $V_{\text{rot}}$, the inclination angle (Inc.), the position angle (PA), the dispersion velocity $\sigma_{\text{gas}}$, the scale height (z0) and the radial velocity ($V_{\text{rad}}$). Here we list the initial values for the fitting process.}
\centering
\begin{tabular}{l|cccc}
\hline\hline
Parameter  & NGC 1512 & NGC 4535 & NGC 7496 \\
\hline
$V_{\text{sys}}$ [\SI{}{\kilo\meter\per\second}](2)& 871.4 & 1953.6 & 1639.2\\
R.A. [deg]  (2) & 60.97557 & 188.58459 & 347.44703\\
Decl. [deg]  (2) & -43.34872 & 8.19797 & -43.42785\\
$V_{\text{rot}} [\SI{}{\kilo\meter\per\second}]$ (1) & 210& 200 & 175 \\
 Inc. [deg] (2) & 42.5& 44.7 & 35.9 \\
 PA [deg] (2) & 261.9& 179.7 & 193.7  \\
 $\sigma_{\text{gas}}$ [\SI{}{\kilo\meter\per\second}] (4) &8&8&8\\
  Z0 [\si{\kilo\parsec}] (3) & 0.68 &0.68  &0.68  \\
 Z0 [arcsec] (3) & 7.5& 8.9 & 7.5\\
 $V_{\text{rad}}$ [\SI{}{\kilo\meter\per\second}]  (4) & 0 & 0 & 0\\
\end{tabular}
\label{tab:Initial Parameters}
 \begin{minipage}{0.95\columnwidth}
        \vspace{1mm}
        {\bf Notes:}(1) \citet{2020ApJ...897..122L}; (2) Galaxy centers taken from \citet{2021ApJS..257...43L} ; (3) \citet{2021ApJ...916...26R}; (4) Default value of \BB. 
    \end{minipage}
\end{table}

Ten free parameters lead to degeneracy in the parameter space, which can cause discontinuities within the results. We deal with this issue by dividing the modeling process into a series of runs, reducing the number of parameters to fit each run.
\begin{itemize}
    \item[-]\textbf{First run}: The set of initial parameters (\autoref{tab:Initial Parameters}) is applied in a first run where all parameters except for $V_{\text{rad}}$ are allowed to vary. Radial velocities are non-dominant for most of the regions in galactic disks and we are interested in pure circular motion, which is why we set radial velocities to zero. By using a weighting of $w(\theta)=|\cos{\theta}|^2$, we made sure that radial motions do not severely influence the fits (compare \autoref{vloss}). The ring width is set to the resolution of the observation (\SI{15}{\arcsecond} for \hi ; $\sim\SI{1}{\arcsecond}$ for CO).
    \item[-]\textbf{Second run}: For the second fitting stage, the central position and systemic velocity of the rings are fixed to the median value of the results for all rings from the first run, which is a common method for galactic modeling \citep[e.g.][]{2017MNRAS.466.4159I,2015PhDT.......216D}. If the rings, especially at large radii, become asymmetric, the fitting process is not reliable anymore which will be seen in a jump in the size of the error bars on all parameters. Rings that have large error bars ($>\SI{\pm30}{\kilo\meter\per\second}$) on the V$_{\text{rot}}$ values are neglected from here on. A flagging of rings after the final fitting run is not advisable since the regularization of the fitting parameters used in the third run might be influenced by badly fitted rings. 
    \item[-]\textbf{Regularization}: The results of the second run can contain numerical scatter, caused by the aforementioned degeneracy in the parameter space. All geometric parameters are further regularized to a function or a constant value to suppress this. The scale height $Z_0$ is regularized by a constant median value. The inclination $i$ and the position angle $\phi$ are regularized to a functional form that represents the fitting results of the second run. To differentiate between real shifts in geometry and numerical scatter we investigated the position velocity diagrams (\autoref{fig:PVs_MAJOR}) and moment maps (\autoref{fig:moment1_HI}) of the observations, which are a good indicator for warped structures. Warps will affect the structure of the major axis position velocity diagram, while radial motions influence the minor axis position velocity diagram \citep[][for a detailed description of this technique]{2021ApJ...923..220D}. All three major axis position velocity diagrams show variations in the velocities which can be an indicator for warps. Therefore we regularize the inclination and position angle by using Bezier functions and do not use more restrictive functional forms.
    \item[-]\textbf{Final run}: Once the geometry of the galaxy is fixed, the kinematics are fitted. The only free parameters in this run are the rotation velocity $V_{\text{rot}}$ and the velocity dispersion $\sigma_{\text{gas}}$. This run yields the final rotation curve. The last two runs are repeated several times to get separate fits for the approaching side, the receding side, and both sides at once.
\end{itemize}
The errors of the parameters for the final model are calculated during the second (for $i$ and $\phi$) and third run (for $V_{\text{rot}}$) based on \BB's\ default error calculation method.

For the CO rotation curves, we use the best-fit results of our \hi\ fits as starting parameters but adapt the fitting algorithm slightly to account for the different properties of CO disks. Two stages are used in this case. The galaxy's systemic velocity and central position are fixed from the beginning to the values found by our \hi models and a $w(\theta)=|\cos{\theta}|^2$ weighting is applied. The scale height $Z_0$ of the disk is expected to be much thinner when observed in CO (compared to $\rm H\,\textsc{i}$). Therefore a thin disk approximation of \SI{100}{\parsec} is assumed (similar to \citet{2022arXiv220316652R} and \citet{2023ApJ...944L..18M}) and held fixed throughout the runs. After a first initial fitting stage where $i$, $\phi$, V$_{\text{rot}}$, and $\sigma_{\text{gas}}$ are kept free to vary, the geometry ($i$ and $\phi$) is fixed to the median value of the best-fit values of all rings. The second run only fits the kinematics $V_{\text{rot}}$ and $V_{\text{disp}}$ for the final model.

   \begin{figure*}[h!]
   \centering
   \includegraphics[width=1\textwidth]{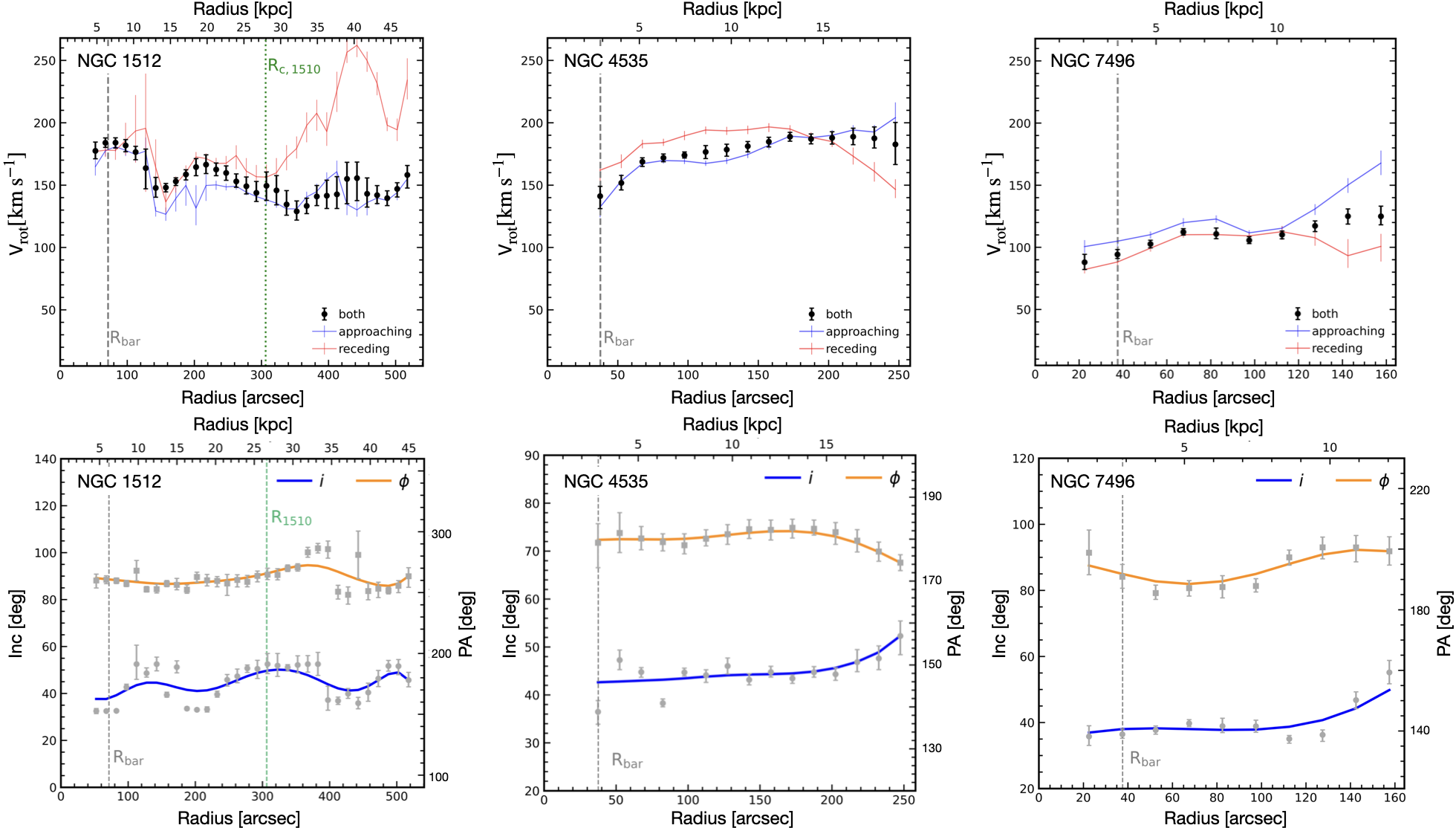}
   \caption{\textit{Top row:} Final \hi\ rotation curves (black circles) for NGC 1512 (left), NGC 4535 (middle) and NGC 7496 (right) overplotted with rotation curves derived by only considering the approaching (blue) and the receding (red) side of the galaxy. The bar radius is indicated with a dashed grey line and the distance of NGC 1510 is indicated with the green dashed line. \textit{Bottom row:} The panels show the adopted geometry in our fitting procedure for NGC 1512 (left), NGC 4535 (middle) and NGC 7496 (right) for the \hi observations. The grey squares and circles represent the inclination $i$ and position angle $\phi$ results for each ring after the second fitting stage, respectively. The blue (inclination) and yellow (position angle) lines represent the regularization function that was used to smooth out the geometry in the final stage of the fitting procedure.}
              \label{fig:ResultsHI}%
    \end{figure*}

\subsection{Radial Velocities}
\label{sec:rad}
For a more complete picture of the atomic gas dynamics, we additionally modeled radial motions inside the galactic disks. For that, the third term of \autoref{vloss} is now included and $V_{\text{rad}}$ is free to vary instead of being fixed to zero as before. As shown in \citet{Cosi}, radial velocities are notoriously difficult to measure in galaxies, since only small changes in their geometry can lead to drastic changes of V$_{\text{rad}}$. The most reliable radial velocities are acquired using the most robust geometric predictions of the galaxies. Therefore, following an approach similar to \citet{2021ApJ...923..220D}, we fit the radial velocities in a singular run using \BB$\,$ with all the other parameters fixed to the previously determined values from our fitting routine.  As mentioned before, radial motions dominantly leave a mark near the minor kinematic axis of galaxies (Section \ref{sec:initial parameters}), which is why for this fitting run a $\sin{\phi}^2$ weighting is applied.

The value of radial velocities derived by \BB\ alone does not inform whether the motion is directed inwards (inflow) or outwards (outflow). The direction of the galaxy's rotation is needed to correctly interpret the nature of the radial motion. By definition of \BB, a positive (negative) $V_{\text{rad}}$ is tracking outflow (inflow) if the galaxy rotates clockwise and inflow (outflow) when it rotates counterclockwise. Assuming that spiral galaxies spin with their arms trailing the rotation movement it is possible to determine the rotation velocities from optical images. We present the results, such that positive values are tracking outflow and negative values inflow. The errors on the radial velocities are determined by \BB's error calculation method. 

\subsection{Mass flow rates}

Detecting radial motions in galaxies can potentially indicate mass transport along the radial direction of galaxies, which is needed to fuel ongoing star formation. To estimate how efficient the transport is, we take the \hi\ surface density profiles $\Sigma_{\text{HI}}\text{(R)}$ into account. Average mass flow rates can be calculated using a simplified approach similar to \citet{Cosi} or \citet{2021ApJ...923..220D}. For that \hi\ intensity profiles $I(R)$ are taken from the \BB$\,$ task ELLPROF, which gives the average flux along an elliptical ring defined by the best-fit geometrical parameters found in our kinematic modeling. For the error of $I(R)$, we adopt the standard deviation provided by the task for each ring. The radial intensity profile can be converted to a face-on \hi\ mass surface density $\Sigma_{\text{HI}}(R)$ when assuming that the \hi\ emission is optically thin:
\begin{equation}
    \frac{\Sigma_{\text{HI}}(R)}{[M_{\odot}\,\text{pc}^{-2}]}=8794\frac{I(R)\cos{i(R)}}{[\text{Jy}\,\text{beam}^{-1}\,\si{\kilo\meter\per\second}]}\left(\frac{\text{B}_{\text{maj}}\text{B}_{\text{min}}}{[\text{arcsec}^2]}\right)^{-1}
\end{equation}
with  B$_{\text{maj, min}}$ being the full width at half maximum of the beam (here \SI{15}{\arcsecond}) and the factor $\cos{i}(R)$ correcting for the inclination of the disk.\\
Using this result, it is now possible to calculate the average \hi\ mass flow $\dot{M}$ for a given radius $R$ by:
\begin{equation}
    \dot{M}_{\text{HI}}(R)=2\pi R\Sigma_{\text{HI}}(R)V_{\text{rad}}(R)
\end{equation}
with $R$ being the radial distance of the ring to the center. The error of $\dot{M}_{\text{HI}}$ is calculated using a Gaussian error propagation of the errors on $\Sigma_{\text{HI}}(R)$ and $V_{\text{rad}}(R)$. Positive values of $\dot{M}_{\text{HI}}$ can be understood as mass outflow and negative values of $\dot{M}_{\text{HI}}$ as mass inflow (accretion). To calculate the total neutral gas mass flow $\dot{M}$, the \hi\ mass flow $\dot{M}_{\text{HI}}$ is multiplied by the factor 1.33 which takes the primordial abundance of helium into account.

\section{Results}
\label{sec:results}
In this section, we present the resulting best-fit parameters derived from our modeling process. First, we will describe the achieved rotation curves from the MeerKAT \hi\ data and then present the results achieved using the PHANGS-ALMA CO data. Values without an error were fixed from the start and not fitted by \BB.

\subsection{HI rotation curves}
\begin{table*}[ht]
\caption{Comparison of best-fit parameters obtained from MeerKAT-\hi and ALMA-CO data. All values in this table are derived from this work. The ring count is given for rings with a thickness of \SI{15}{\arcsecond} width for \hi\ and \SI{1}{\arcsecond} for CO.}
\centering
\begin{tabular}{l|cc|cc|cc}

 & \multicolumn{2}{c}{NGC 1512} \vline & \multicolumn{2}{c}{NGC 4535} \vline & \multicolumn{2}{c}{NGC 7496} \\

 & \hi & CO & \hi & CO & \hi & CO\\ \hline\hline
$V_{\text{sys}}$ [\SI{}{\kilo\meter\per\second}] & $880.8\pm6.3$ & $880.8$ & $1951.7\pm7.7$ & $1951.7$  & $1639.2\pm5.2$ & $1639.2$ \\ 
R.A. [deg] & 60.97557 & 60.97557 & 188.58459 & 188.58459 & 347.44703 & 347.44703\\
Decl. [deg] & -43.34872 & -43.34872 & 8.19797 & 8.19797 & -43.42785 & -43.42785\\  
$z_0$ [\SI{}{\kilo\parsec}] & $1.06\pm0.08$ & 0.1& $0.79\pm0.08$ & 0.1 & $0.83\pm0.1$ & 0.1  \\
$i$ median [\SI{}{\degree}] & $43.47\pm5.4$ & $43.3\pm3.03$& $44.3\pm2.93$ & $43.7\pm4.5$ & $37.99\pm3.59$  & $44.49\pm2.72$  \\
$\phi$ median [\SI{}{\degree}] & $265.8\pm10.8$ & $252.53\pm9.15$& $179.91\pm5.75$ & $182.77\pm7.78$ & $191.8\pm7.22$  & $201.22\pm7.17$  \\
V$_{\text{disp.}}$ [\SI{}{\kilo\meter\per\second}]& $4.28\pm4.23$ & $5.97\pm2.33$ & $6.53\pm4.52$& $3.88\pm2.02$ &  $7.32\pm3.98$& $5.23\pm2.54$ \\
V$_\text{rot, max}$ [\SI{}{\kilo\meter\per\second}]& $177.3\pm16.4$  & $204.8\pm9.7$& $188.8\pm6.2$ & $181.93\pm6.09$ & $125\pm5.8$  & $118.11\pm6.09$ \\
R$_{\text{max}}$ [kpc] & 517.5 & 89.93 & 247.5 & 109.33 & 157.5 & 74.54 \\
Ring Count & 32 & 36 & 15 & 55 & 10 & 45 \\
\end{tabular}
\label{tab:HI-CO-results}
\end{table*}
\autoref{fig:ResultsHI} shows the best-fit HI rotation velocities, as well as the results achieved using one side (receding or approaching) of the galactic disk. The \hi\ rotation curves reach velocities between ${\sim}90$ and ${\sim}\SI{200}{\kilo\meter\per\second}$. Successful fits were achieved for 32 rings of $\SI{15}{\arcsecond}$ width for NGC 1512, 15 rings for NGC 4535, and 10 rings for NGC 7496. The innermost rings of the RCs exhibit velocities close to the maximum of the curves. A steep rise of the velocities is therefore expected in the central regions, which are not covered by the \hi\ observations. 

The approaching and receding rotation curves are typically close to each other (within $\approx\SI{20}{\kilo\meter\per\second}$) for most radii in all three galaxies. Differences in the approaching and receding fitting results are caused by asymmetries in the disk. Especially for NGC 1512, we see very pronounced differences between the approaching and receding sides at radii larger than \SI{300}{\arcsecond}, which are caused by the influence of the companion galaxy NGC 1510. 

Our best-fit values of the central positions and systemic velocities are provided in \autoref{tab:HI-CO-results}. We took the median for these two parameters for all successfully fitted rings after the first stage and used these values for the following stages. The galactic centers agree well (same pixel) with the previously determined values (our initial parameters). For the systemic velocity, the values were in very good agreement with the previously found values for NGC 4535 and NGC 7496. The new value for NGC 1512 of $880.8\pm6.3\,$\SI{}{\kilo\meter\per\second} is still within the errors of the previous $871.4\pm5\,$\SI{}{\kilo\meter\per\second}.

The inferred disk geometry (\autoref{fig:ResultsHI} bottom row) for all three galaxies warps at large galactocentric radii that entail subtle changes in $\phi$ and $i$ moving outward. For NGC 4535 and NGC 7496, a warp in inclination at large radii and evolving position angles throughout the disks are modeled. These warps are also indicated by the deviation from constant velocities at similar radii in the major axis position velocity diagrams (\autoref{fig:PVs_MAJOR}). NGC 1512 exhibits the most complex geometry with a varying behavior of the inclination at ${\sim}\SI{40}{\degree}$ throughout its radial extent. Such oscillations are also evident in the position velocity diagram of the major axis and are likely caused by the presence of the companion galaxy NGC 1510.

   \begin{figure*}[h!]
   \centering
   \includegraphics[width=1\textwidth]{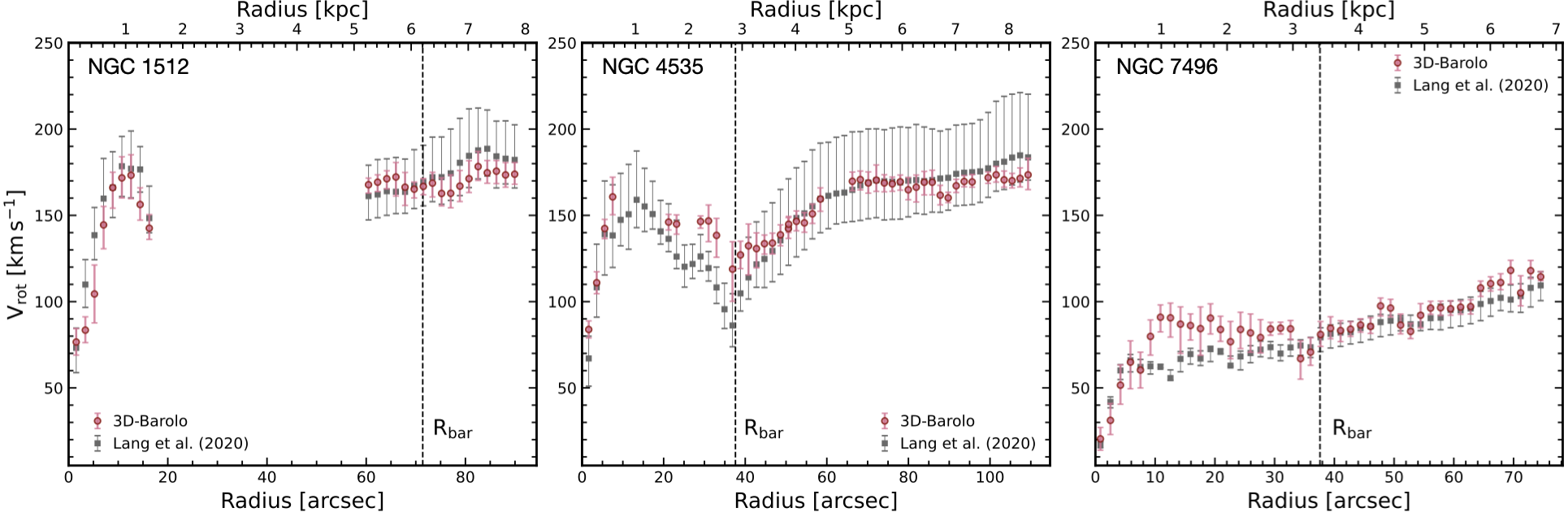}
   \caption{PHANGS-ALMA CO rotation curve of the sample produced by 3D-Barolo (black circles). The previous fits by \citet{2020ApJ...897..122L} are plotted as grey squares. The radius of the galactic bar is indicated by the vertical dashed line. Some bins of the 3D-Barolo rotation curves did not converge and are therefore not represented. The two different methods lead to very similar results overall. The rotation curve of NGC 1512 has a large area without data points, which corresponds to regions with too few emissions for reliable fitting.}
              \label{figs:COresults}%
    \end{figure*}
\subsection{CO Rotation Curves}
The circular velocities derived from the CO fitting procedure are presented in \autoref{figs:COresults} in red. We overplotted the results with previous RCs derived by \citet{2020ApJ...897..122L}, which represent the latest kinematic analysis of these galaxies. Their method differs from ours as they assumed that the CO emitting disks are infinitely thin and rotating with a single, global orientation and thus derived V$_{\text{sys}}$, $i$, PA, and V$_{\text{rot}}$ with 2D tilted ring fitting (with $\phi$, $i$ and V$_{\text{sys}}$ held fixed). For a more direct comparison, we adopted the same radial bins as \citet{2020ApJ...897..122L}.

Our RCs based on CO recover the central rising part of the curves well. For all three galaxies a strong rise within the first $\approx\SI{1}{\kilo\parsec}$ can be seen with a flatter behavior for larger radii. The rotation curve of NGC 1512 has a large area without data points, which correspond to regions with too few emission for reliable fitting. NGC 4535 shows a pronounced dip around the bar radius, then rises again and flattens out at around \SI{6}{\kilo\parsec}. NGC 7496 shows overall lower circular velocities than the other two galaxies and rises slowly until the largest radii.

The adopted best-fit parameters are listed in \autoref{tab:HI-CO-results}. Overall the circular velocities derived by \BB~ agree well with the values from the previous work by \citet{2020ApJ...897..122L} (compare \autoref{figs:COresults}). The best-fit parameters of this work (\autoref{tab:HI-CO-results}) are also in good agreement with the previously derived values (our starting parameters \autoref{tab:Initial Parameters}). For NGC 7496, our three-dimensional CO fit produces a higher inclination and position angle. While the errors of both values are overlapping, we still investigated if the new methodology is causing the different results in the geometry. Different values for the scale height for \hi\ did not have a big impact on our fitting results (1pc and 500pc stay within the errors of the final fit; see Appendix \ref{sec:scale}) and therefore we trust that they do not have a decisive impact on our CO fits. Additionally, we fitted the galaxy by fixing the geometry to the starting parameters, but this produced unrealistic rotation curves with large jumps in velocity ($\pm\SI{20}{\kilo\meter\per\second}$ between many bins). Because of the better agreement with the \hi results we think that the geometric values from \citet{2020ApJ...897..122L} are more trustworthy for the inner CO disk of NGC 7496.

\section{Discussion}
\label{discussion}

   \begin{figure*}[h!]
   \centering
   \includegraphics[width=0.85\textwidth]{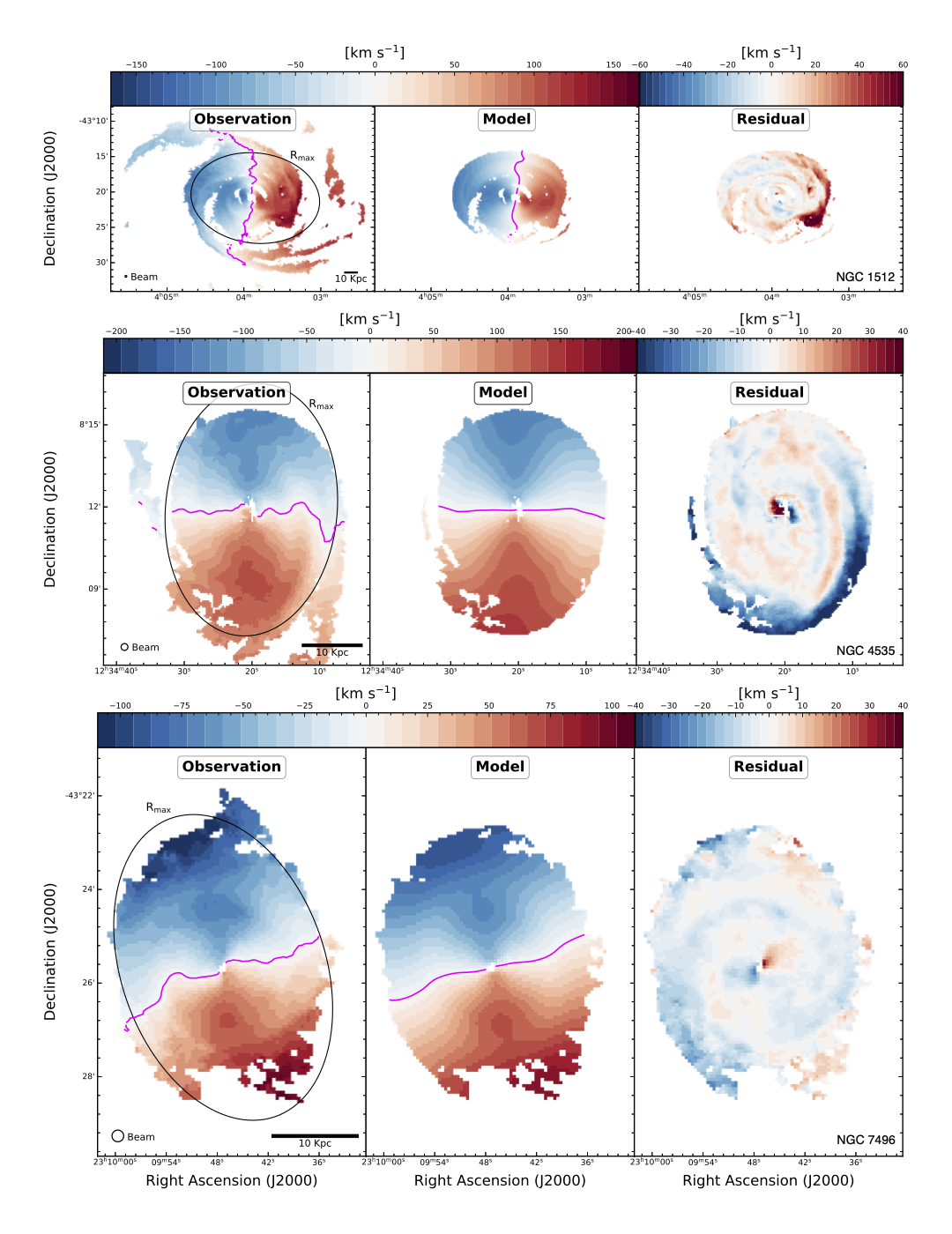}
   \caption{\textit{Left column:} Moment 1 map of the \hi\ observations for NGC 1512 (top), NGC 4535 (middle), and NGC 7496 (bottom). The black ring indicates the outermost radius that was used for the model. The pink line tracks the systemic velocity line. \textit{Middle column}: Moment 1 map of the \hi\ model using the best-fit parameters found. \textit{Right column}: Residual of the observation and the model. Models and residuals larger than R$_{\text{max}}$ are not shown due to unphysical results as discussed in Section \ref{sec:Methods} as part of the second run.}
              \label{figs:Residuals}%
   \end{figure*}

   \begin{figure*}[h!]
   \centering
   \includegraphics[width=0.85\textwidth]{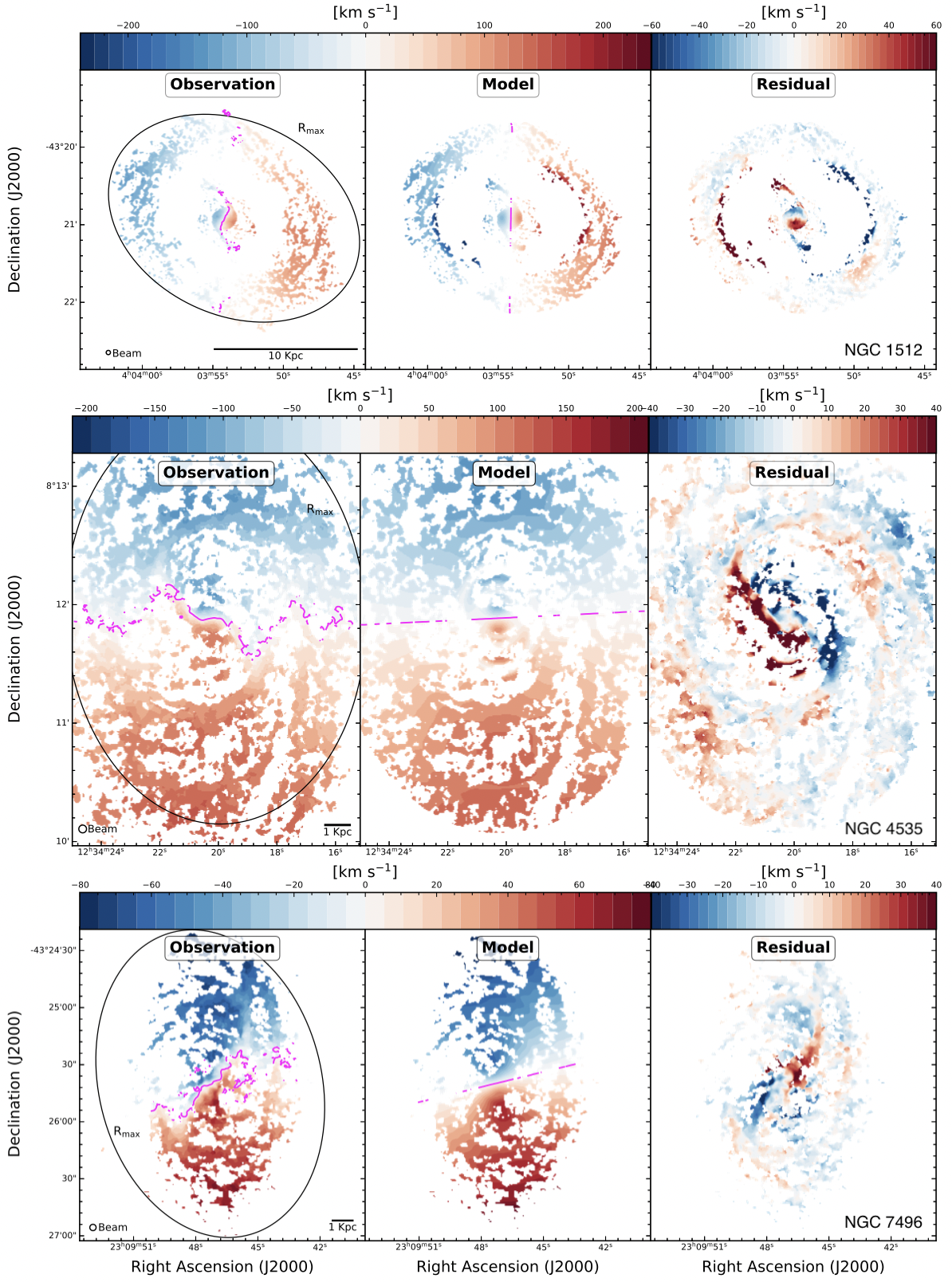}
   \caption{\textit{Left column:} First-moment map of the CO observations for NGC 1512 (top), NGC 4535 (middle), and NGC 7496 (bottom). The black ring indicates the outermost radius that was used for the model. The pink line tracks the systemic velocity line. \textit{Middle column}: First-moment map of the \hi\ model using the best-fit parameters found. \textit{Right column}: Residual velocity field between the observation and the model. Models and residuals larger than R$_{\text{max}}$ are not shown due to unphysical results as discussed in Section \ref{sec:Methods} as part of the second run.}
              \label{figs:CO_Residuals}%
   \end{figure*}
   
In this section, we discuss the reliability and implications of previously presented results and evaluate how our assumptions affect them.
\subsection{Reliability of the models}
The \hi\ rotation curves (\autoref{fig:ResultsHI}) exhibit a slight rise at the inner most radii, but overall \hi\ seems to track mostly the flat regime for all three galaxies. The velocities in this regime correspond to very typical values compared to previous studies. Kinematic \hi\ studies on larger samples such as THINGS \citep{2008AJ....136.2648D} or more recent publications \citep{2021ApJ...923..220D} establish a typical range of \SI{100}{\kilo\meter\per\second} to \SI{250}{\kilo\meter\per\second}. NGC 4535 and NGC 1512 therefore show normal velocity behaviour with NGC 7496 being at the lower end of the typical range. \citet{2021ApJ...923..220D} derived a \hi-based rotation curve for NGC 4535 using \BB, finding a lower inclination angle (\SI{36}{\degree} compared to our \SI{44.3}{\degree}) resulting in higher rotation velocities. The velocities are especially deviating at very large radii because of the assumed constant inclination in contrast to the slightly warped structure of our model. Applying a Tully-Fisher relation similar to \citet{2015ApJ...802...18M} implies a velocity of about \SI{182}{\kilo\meter\per\second} which is also in good agreement with our rotation curve. The rotation curve for NGC 1512 published by \citet{2009MNRAS.400.1749K} shows higher velocities overall, as they assume a constantly increasing inclination in contrast to our varying inclination angle. The better resolution of the MeerKAT data used for our RC (\SI{15}{\arcsecond} compared to their \SI{65}{\arcsecond}) reveals a more detailed picture of the geometry, which can explain the differences.

A direct comparison between the observation and the models can help to understand how the assumptions affect our fits. For that, we created residual maps by subtracting the modeled velocities from the observations (\autoref{figs:Residuals} for \hi\ and \autoref{figs:CO_Residuals} for CO). For NGC 1512 we see overall residual velocities between \SI{-20}{\kilo\meter\per\second} and \SI{20}{\kilo\meter\per\second} in \hi\ throughout most of the disk, but it has a large spike on the receding side of the disk at the location of NGC 1510. For NGC 4535 and NGC 7496, the residual velocities are also between \SI{-20}{\kilo\meter\per\second} and \SI{20}{\kilo\meter\per\second} in \hi\ for most of the disk, which supports our modeling routine. For radii $\lesssim R_{\text{Bar}}$, large residual velocities of up to \SI{60}{\kilo\meter\per\second} are present and are likely caused by non-circular motions associated with the bar
 \citep{2016MNRAS.457.2642S}. These non-circular motions are not part of our model. Both galaxies further show spiral patterns in the residual maps of \hi\ that correspond to the position of their spiral arms implying streaming motions along the spiral arms. The model tries to account for real radial motions within the observation, leading to an overestimation of regions without radial motions and vice versa.

\subsection{Caveats}
Several effects can influence the outcome of a tilted ring fit. The tilted ring approach relies on assumptions and approximations that might not always be fully appropriate for real galaxies. Models are always symmetric, while real \hi disks are known to be subject to asymmetries \citep{2005A&A...442..455E,Cosi}. The one-sided fitting results (\autoref{fig:ResultsHI}) in our sample all show asymmetries, especially at large radii. Modeling the kinematics of interacting galaxies (like NGC 1512 and NGC1510) is therefore challenging and not reliable due to the distortion introduced by tidal torques.

While three-dimensional modeling does lessen the effects of beam smearing \citep{2015ApJ...802..125R}, it also comes with its own drawbacks. As described by \citet{2022arXiv220316652R}, when the model emission of a given ring is compared to the corresponding region on the sky, it is assumed that the emission comes from a fixed radial interval within the disk. This would almost be true if the rings were razor-thin vertically or two-dimensional. For rings that have a volume, emission from above and below the mid-plane is taken into account, which results in overlapping emission from other radii that contaminate a given interval. This effect is boosted in particular for moderately to highly inclined galaxies. Therefore, an observation from one given location might be used for setting the parameters of several rings. 

In reality, the outer gas disk of galaxies is often flared, which means that its scale height increases with radius \citep{2021A&A...655A.101D}. In the model, the average disk thickness of a large galaxy sample was used as an initial parameter and the gas disk was fixed to have a constant scale height after a first fitting run. It is therefore expected that the disks of the three galaxies are generally thinner in the central regions and can be thicker in the outer regions (depending on how strongly the \hi\ is flared) than the assumed fixed scale height parameter. We tested how this affects the final parameters by varying the fixed scale height of our models from unrealistically small ($Z_0=\SI{1}{\parsec}$) to unrealistically large ($Z_0=\SI{2}{\kilo\parsec}$) constant values over several fitting runs. The outcome of these fits is presented in \autoref{sec:scale}. Most of the results are still within the error range of our final adopted model. Assuming a constant average scale height does not seem to be a dominant source of uncertainty.

\subsection{Combined fitting of \hi and CO data}

\begin{figure*}[h!]
   \centering
   \includegraphics[width=1\textwidth]{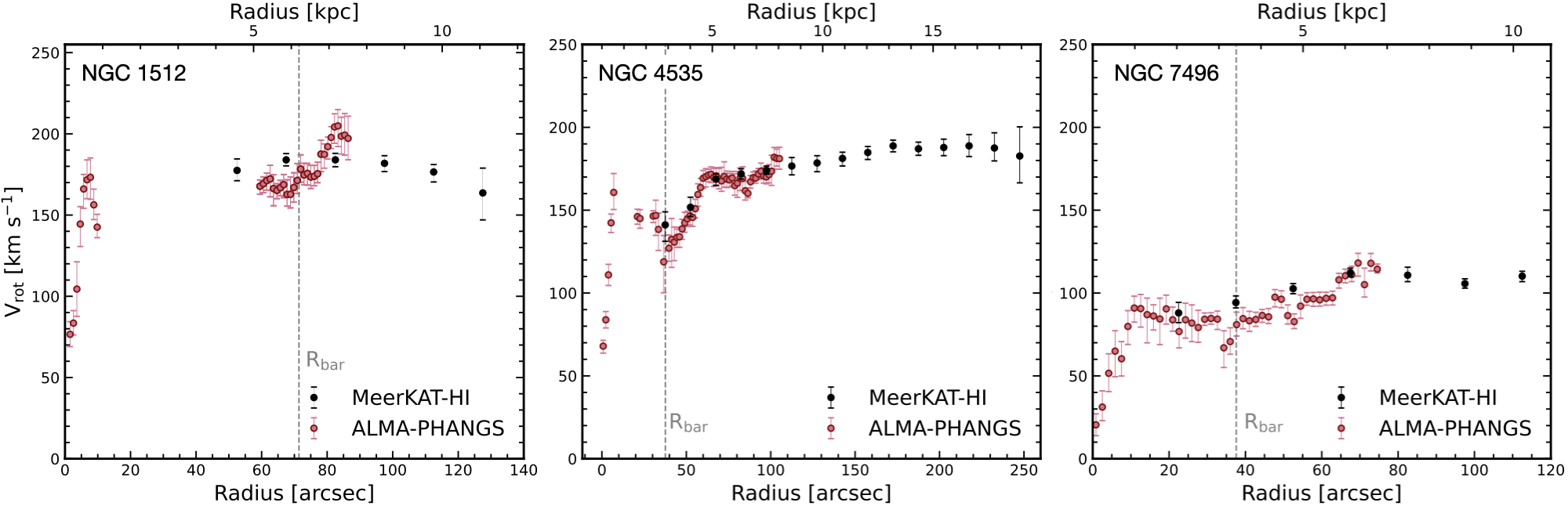}
   \caption{MeerKAT-\hi\ rotation curves (black circles) are directly compared to PHANGS-ALMA CO rotation curves (pink circles) derived in this work.}
              \label{figs:CO-HI}%
    \end{figure*}

We compared the circular velocities of the different tracers in \autoref{figs:CO-HI}. Combining the two, leads to a more complete view of the disk dynamics, covering the inner regions by CO to the most extended radii probed by \hi. The \hi\ and CO circular velocities are in agreement for most bins where measurements overlap. For all three galaxies, we find an offset towards higher velocities in \hi\ at a radius corresponding to the scale of the galactic bar. Especially in NGC 1512, the velocities derived from the two tracers seem to track different motions at the largest radii recovered by CO. Overall, the CO rotation curves draw a more granular and less smooth picture of the kinematics, which can be partially attributed to the higher resolution in CO. \citet{2014ApJ...784....4C} and \citet{2016AJ....151...94F} reported similar local differences between the CO and \hi\ rotation velocities in M51. CO is associated with tracing denser cold gas (compared to \hi), while emission from atomic gas is more broadly distributed throughout the whole disk. Bars or spirals cause more streaming motions in comparison to the inter-arm regions which will contribute to the observed velocities. Additionally, CO emission arises from a thinner disk compared to \hi, which causes CO to trace the gravitational potential differently than \hi\ \citep[e.g.][]{2018ApJ...860...92L, 2019ApJ...882...84L, 2015ApJ...799...61Z}.

The residuals of the CO velocity fields reveal that assuming pure circular velocities is a very simplified way to model the complex dynamics in the inner disk. These regions are heavily influenced by non-circular motions. Large differences especially in the systemic velocity line (pink) can be seen when comparing observations and models of the first moments. The larger residuals of the CO models in comparison to \hi\ imply that the parameters determined from \hi\ are less disturbed by non-axisymmetric influences that are in general difficult to model correctly. The inner CO disk does not necessarily share the same orientation as the outer parts of the galaxy, but deriving first orientations for the \hi disk and then using these \hi-based orientations as initial parameters for the CO observations worked well for our observations (see Section \ref{sec:results} and \autoref{figs:CO-HI}). This is why we recommend using \hi to correctly model the disk orientation parameters before deriving CO-based RCs.

\subsection{Radial motions}

\label{radial}
   \begin{figure*}[ht]
   \centering
   \includegraphics[width=1\textwidth]{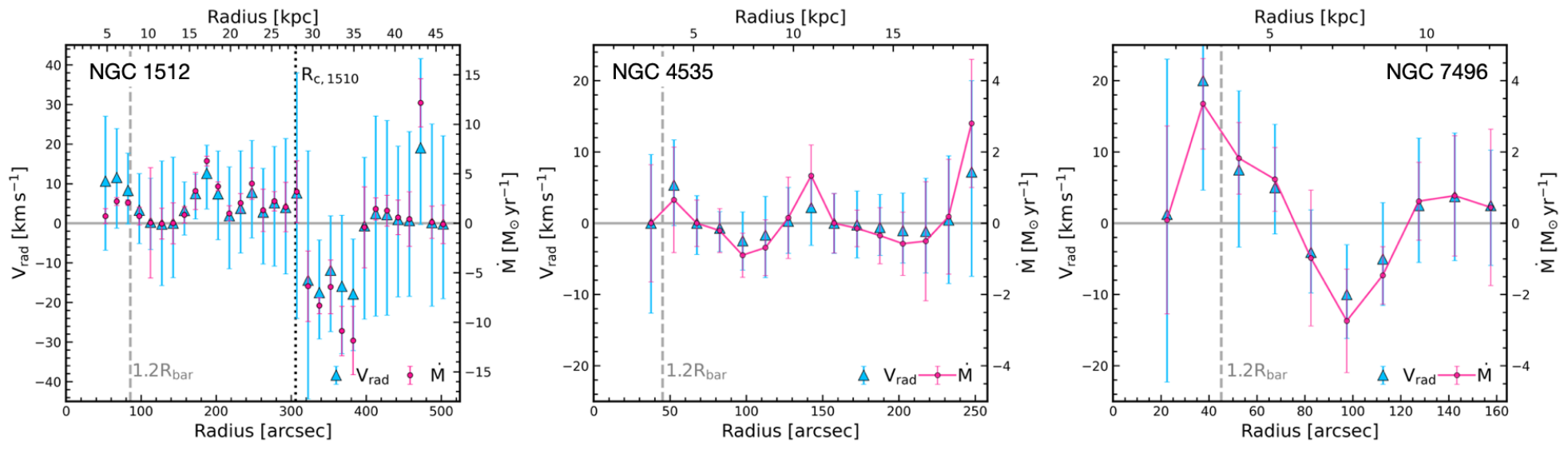}
   \caption{Radial velocities of HI determined by the tilted ring model on the MeerKAT-\hi\ observations (blue triangles). Mass flow rates show the neutral atomic gas (purple circles). The bar radii and the radial position of NGC 1510's center are indicated with the dashed and dotted lines.}
              \label{figs:Vrad}%
   \end{figure*}

In \autoref{figs:Vrad} we show the radial motions that were extracted from the tilted ring model (see Section \ref{sec:rad}). All galaxies show radial in- and outflow velocities up to $\sim|\SI{20}{\kilo\meter\per\second}|$. In all of the three galaxies, we note that the bar affects the direction and magnitude of the radial motions strongly. We also show the mass flow rates $\dot{M}$ indicated as purple markers in \autoref{figs:Vrad}. The general direction of the mass flow traces the direction of the radial velocities, but the magnitude differs because of its additional dependence on how much gas is actually in that radial bin (traced by $\Sigma_{\text{HI}}$). 

The values for NGC 4535 and NGC 7496 vary between inflow and outflow up to -2 and 2 M$_{\odot}$yr$^{-1}$. Stronger mass flow rates can be observed in NGC 1512 where values of $\dot{M}>10\,{\text{M}}_{\odot}\,\text{yr}^{-1}$ are reached. We see the strongest changes in NGC 1512 at around $R\approx\SI{300}{\arcsecond}$, where the companion galaxy NGC 1510 is located. At smaller radii, NGC 1512 shows a dominant outflow with velocities of less than \SI{10}{\kilo\meter\per\second} while it suddenly changes to strong inflow at 300 to \SI{400}{\arcsecond} with a magnitude of about \SI{20}{\kilo\meter\per\second}. These shifts could be caused by the gravitational influence both galaxies have on each other. It is more likely that the fitting tries to account for the large residuals caused by the presence of NGC 1510 and might therefore overestimate the radial velocities at this point. This uncertainty makes the radial bins larger than $\text{R}_{c, 1512}$ (Radial distance of NGC 1510 to the center of NGC 1512) unreliable and we further ignore them in the ongoing analysis.

Existing studies of radial gas flow from HI observations of local galaxies \citep[e.g.,][]{2008AJ....136.2720T, 2014ApJ...784....4C, 2016MNRAS.457.2642S, Cosi} have mostly derived radial velocities using 2D harmonic decomposition.  In this work, in contrast, we employ a 3D modeling approach to gather information about the radial movement of gas.

Overall, the magnitude of $V_{\text{rad}}$ and $\dot{M}$ are in agreement with typical values for comparable galaxies. \citet{2021ApJ...923..220D} recently published radial motions and mass flow rates for NGC 4535 using \BB. The magnitude of the mass flow is very similar to the results in this work, but the orientation of the flow differs. Deriving radial velocities based on tilted ring models is very sensitive to the assumed geometrical parameters  \citep[e.g.][]{Cosi}. Relatively small differences in inclination can lead to a complete flip of the inferred flow directions. Therefore, the differences in the results, even though the same methods were used, are likely to be caused by the differently assumed inclination angle ($i=\SI{43.7\pm4.5}{\degree}$ with a slight rise at large radii in this work compared to $i=\SI{36}{\degree}$).

\begin{table}[ht]
\caption{Star formation rates taken from \citet{2019ApJS..244...24L} compared to the average radial velocities and mass flow rates derived in this paper. Negative (positive) values represent inflow (outflow).}
\centering
\begin{tabular}{lccc}
\hline\hline
& SFR & $\langle V_{\text{rad}}\rangle$ & $\langle\dot M\rangle$ \\
&  [M$_{\odot}\,\text{yr}^{-1}]$   & [\si{\kilo\meter\per\second}] & [M$_{\odot}\,\text{yr}^{-1}]$\\
\hline
NGC 1512 & $0.75\pm0.75$  &  $4.47\pm8.1$ & $2.02\pm1.85$\\
NGC 4535 & $2.04\pm1.6$ & $-0.55\pm5.57$ & $-0.13\pm1.79$ \\
NGC 7496 & $2.09\pm1.6$ & $-0.27\pm5.39$ & $-0.03\pm1.49$\\
\hline
\end{tabular}
\label{tab:SFRs}
\end{table}

We aim to know whether radial motions in neutral atomic gas of our three galaxies can sustain current star formation rates (SFRs). Table \ref{tab:SFRs} compares the star formation rates taken from \citet{2019ApJS..244...24L} to our derived average radial velocities and average mass flow rates. The tilted ring model is susceptible to underlying non-axisymmetric dynamical features (bars or spirals), which are very difficult to incorporate and mainly influence the inner regions of galactic disks. To prevent a strong influence of this effect we calculated an average in radial velocity and mass flow rates only for radii larger than $1.2\,\text{R}_{\text{bar}}$, where the influence of the bars is almost negligible \citep{1998AJ....116.2136A}. 

The averaged mass flow rates and their uncertainties for NGC 4535 and NGC 7496 are largely consistent with no mass flow at all. The average mass flow in NGC 1512 is directed radially outwards, which might again be a sign of the gravitational influence of its companion NGC 1510. The limited sample size and the presence of substantial uncertainties in our data preclude us from making definitive conclusions. However, other studies, such as \citet{2021ApJ...923..220D} and \citet{2014ApJ...780..105R}, have also indicated that a significant portion of the observed SFRs cannot be ascribed to radial accretion of neutral atomic gas. Such results suggest that observed SFRs may not solely stem from the radial accretion of neutral atomic gas. Additional factors like high-velocity clouds, episodic accretion, or gravitational interactions could contribute \citep{2016A&A...589A.120K, 2014ApJ...780..105R}. This study provides a methodological framework for future investigations when a larger sample of similar observations is available despite the need for a cautious interpretation of our specific results.

\section{Conclusions}

In conclusion, our study delved into the dynamics of both the neutral atomic and CO-traced molecular components within the galactic disks of the three nearby spiral galaxies NGC 1512, NGC 4535, and NGC 7496. Using MeerKAT \hi\ and PHANGS-ALMA CO observations at resolutions of \SI{15}{\arcsecond} and $\sim\SI{1}{\arcsecond}$, respectively, we employed the \BB$,$ tilted ring fitting code to derive robust tilted ring models. Our key findings and insights include:
\begin{enumerate}
    \item We established reliable \hi\ tilted ring models and utilized them to constrain the geometry of the CO observations, deriving corresponding tilted ring models for molecular components.

    \item Signatures of warps were identified in all three galactic disks through position-velocity diagrams and kinematic fitting results. NGC 4535 and NGC 7496 exhibited slight warps at large \hi\ radii, while NGC 1512 displayed a more intricate structure, likely influenced by the proximity of NGC 1510.

     \item Comparison with the approach used in \citet{2020ApJ...897..122L} affirmed the validity of our approach. Similar orientations and dynamics in both analyses provided confidence in the robustness of our assumptions.

     \item Rotation velocities based on neutral atomic and molecular gas were combined and compared. The agreement between the two differently traced rotation curves was observed for most overlapping radii, demonstrating the effectiveness of combining these tracers to derive RCs across the entire disk.

     \item Investigation into radial motions and mass flow rates revealed minimal mass flow in all galaxies, with a small outflow in NGC 1512 potentially attributed to the presence of NGC 1510. The methodology employed for deriving radial motions and mass flow rates serves as a viable approach for future studies when expanded datasets become available.

\end{enumerate}

With the ongoing \hi coverage of the PHANGS-ALMA sample, it will be possible to apply the approach presented here to the whole sample. With a larger number of galaxies, we will be able to gain new insights and formulate reliable conclusions on how the dynamics and the interplay of \hi and molecular gas determine the structure of spiral galaxies. It will be possible to investigate the fueling of star formation via radial infall of neutral atomic gas similar to the approach presented in this paper and in \citet{2021ApJ...923..220D}. With the possibilities of fitting the scale height, it would be interesting to investigate how and if it is related to the rotation velocity, as indications for such a relation were already found by \citet{2010A&A...515A..62O}.

\begin{acknowledgements}
The work of AKL is partially supported by the National Science Foundation under Grants No. 1615105, 1615109, and 1653300. \\
ES acknowledges funding from the European Research Council (ERC) under the European Union’s Horizon 2020 research and innovation programme (grant agreement No. 694343). \\
MQ acknowledges support from the Spanish grant PID2019-106027GA-C44, funded by MCIN/AEI/10.13039/501100011033. \\
SKS acknowledges financial support from the German Research Foundation (DFG) via Sino-German research grant SCHI 536/11-1.\\
ER acknowledges the support of the Natural Sciences and Engineering Research Council of Canada (NSERC), funding reference number RGPIN-2022-03499.\\
RCL acknowledges support provided by a National Science Foundation (NSF) Astronomy and Astrophysics Postdoctoral Fellowship under award AST-2102625.\\
RSK acknowledges financial support from the German Research Foundation (DFG) via the collaborative research center (SFB 881, Project-ID 138713538) “The Milky Way System” (subprojects A1, B1, B2, and B8). He also thanks for funding from the Heidelberg Cluster of Excellence ``STRUCTURES'' in the framework of Germany’s Excellence Strategy (grant EXC-2181/1, Project-ID 390900948) and for funding from the European Research Council via the ERC Synergy Grant ``ECOGAL'' (grant 855130). \\
EWK acknowledges support from the Smithsonian Institution as a Submillimeter Array (SMA) Fellow and the Natural Sciences and Engineering Research Council of Canada. \\
This work has received funding from the European Research Council
(ERC) under the European Union’s Horizon 2020 research and innovation
programme (grant agreement No. 882793 ``MeerGas'').
LN acknowledges funding from the Deutsche Forschungsgemeinschaft (DFG, German Research Foundation) - 516405419.\\
HAP acknowledges support by the National Science and Technology Council of Taiwan under grant 110-2112-M-032-020-MY3.\\
TGW acknowledges funding from the European Research Council (ERC) under the European Union’s Horizon 2020 research and innovation programme (grant agreement No. 694343).

\end{acknowledgements}

\bibliographystyle{aa} 
\bibliography{references}

\begin{appendix} 

\section{Influence of the scale height parameter}
Since the true height of the disks is not known for our sample we made assumptions regarding the scale height parameter. We took a simple approach, by using a constant value for all radii. The initial parameter is based on an average value for \hi\ scale heights by \citet{2021ApJ...916...26R}, based on a large \hi\ survey. By using a constant value we cannot take any variations of scale height into account, although large \hi\ disks are known to flare at large radii \citep{2021A&A...655A.101D, 2021ApJ...916...26R}. 

To justify the simple approach, we were interested in how much the scale height affects our final rotation curves. We conducted eight additional fitting runs, which fixed the constant scale height parameter to a variety of values. All other parameters were left unchanged from our original best-fit model. The scale heights range from unrealistically small (\SI{1}{\parsec}) to unrealistically large (\SI{2000}{\parsec}) values.

\begin{figure*}[h!]
   \centering
   \includegraphics[width=1\textwidth]{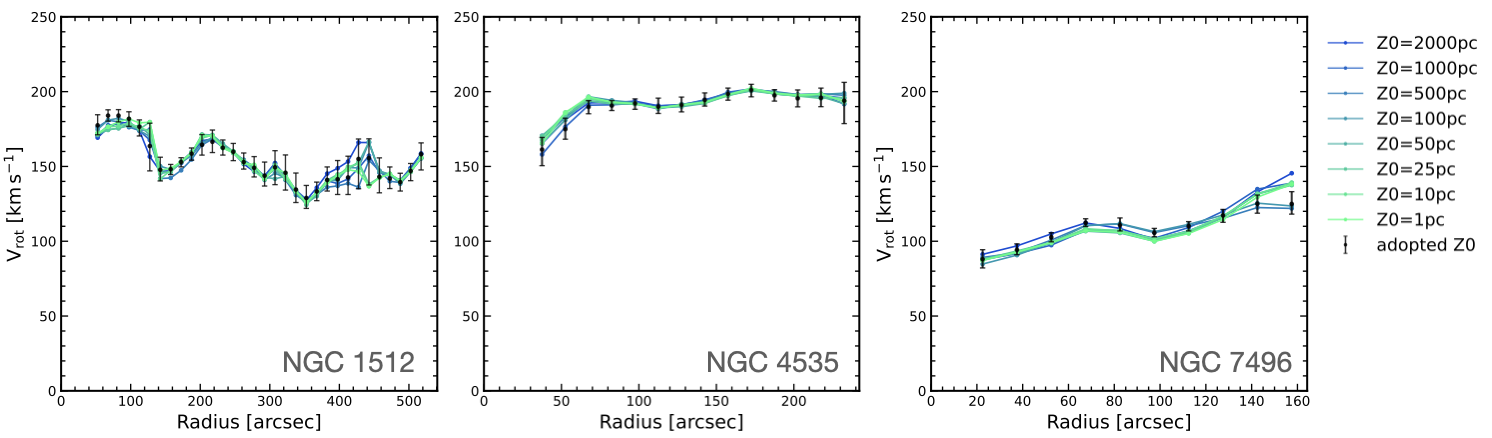}
   \caption{Fitting results achieved for varying constant values of scale height for 9 different values for NGC 1512 (left), NGC 4535 (middle), and NGC 7496 (right). The scale height was chosen from unrealistically small (green) to unrealistically large (blue) values. The final rotation curve adopted in this paper is plotted in black. }
              \label{figs:scale}%
   \end{figure*}

The results of this procedure are presented in \autoref{figs:scale}. Most of the RCs obtained in the additional runs are within the errors of our best-fit model (black circles), which supports the simple approach we have used. A trend to how the thickness of the disk affects the rotation velocities cannot be identified.

\label{sec:scale}

\end{appendix}
\end{document}